\documentclass[fleqn,10pt]{SelfArx} 
\usepackage[table]{xcolor}
\usepackage[authoryear]{natbib}
\definecolor{color1}{RGB}{0,0,90} 
\definecolor{color2}{RGB}{0,20,20} 
\usepackage{hyperref} 
\hypersetup{hidelinks,colorlinks,breaklinks=true,urlcolor=color2,citecolor=color1,linkcolor=color1,bookmarksopen=false,pdftitle={Title},pdfauthor={Author}}
\definecolor{Gray}{gray}{0.9}
\definecolor{LGray}{gray}{0.95}

\usepackage{babel}
\usepackage[utf8]{inputenc}
\usepackage[T1]{fontenc}
\usepackage{mathptmx}       
\usepackage{helvet}         
\usepackage{courier}        
\usepackage{type1cm}        
\usepackage{makeidx}         
\usepackage{graphicx}        
\graphicspath{{Figures/}}
\usepackage{multicol}        
\usepackage[bottom]{footmisc}
\usepackage{amssymb}

\usepackage{amsthm}
\usepackage{url}

\usepackage{array}

\renewcommand{\emph}[1]{\textbf{#1}}

\theoremstyle{definition}

\graphicspath{ {./Figures/} }
\DeclareMathAlphabet{\mathcal}{OMS}{cmsy}{m}{n}
\SetMathAlphabet{\mathcal}{bold}{OMS}{cmsy}{b}{n}

\makeindex             

\JournalInfo{Cryptocurrency Network Analysis} 
\Archive{Chapter} 

\PaperTitle{Cryptocurrency Network Analysis}
\Authors{Natkamon Tovanich\textsuperscript{1}, Célestin Coquidé\textsuperscript{2}, Rémy Cazabet\textsuperscript{2}}

\affiliation{\textsuperscript{1} CREST, CNRS, École Polytechnique, Institute Polytechnique de Paris, 91120 Palaiseau, France
}
\affiliation{\textsuperscript{2} Universite Claude Bernard Lyon 1, CNRS, INSA Lyon, LIRIS, UMR5205, F-69622 Villeurbanne, France}

\newcommand{\keywordname}{Keywords} 

\Abstract{\textbf{Definition.} Cryptocurrency network analysis consists of applying the tools and methods of social network analysis to transactional data issued from cryptocurrencies. The main difference with most online social networks is that users do not exchange textual content but instead value —in systems designed mainly as cryptocurrency, such as Bitcoin— or digital items and services in more permissive systems based on smart contracts such as Ethereum.}

\begin{document}
\flushbottom
\maketitle 

\renewcommand{\contentsname}{Table of Contents}
\renewcommand{\abstractname}{}
\flushbottom 

\tableofcontents 

\thispagestyle{empty} 

\section{Introduction}

Online social network platforms such as Facebook or Twitter have transformed the field of social network analysis by allowing access to large amounts of data on social interaction between individuals. Cryptocurrency represents a new step in this direction by allowing researchers and analysts to access large amounts of data about economic transactions between entities. Both types of data share many similarities, including their natural representation as networks, large scale, dynamic nature, and influence by and upon society as a whole.

\section{Key Points}

\begin{itemize}
    \item Cryptocurrency transactional data can be accessed and collected freely, providing data similar in nature to online social network interactions.
    \item Bitcoin and Ethereum, the most popular cryptocurrencies as of 2024, are particularly relevant to this type of analysis.
    \item Many research questions have been explored on these networks, some of the most popular being user de-anonymization, illegal activity and fraudulent user identification, trading strategy analysis, network properties analysis, and price forecasting.
\end{itemize}

\section{History and Technical Premises
}
The history of cryptocurrencies begins with Bitcoin, introduced in 2008 by an individual or group using the pseudonym Satoshi Nakamoto. The white paper \cite{nakamoto2008bitcoin} outlines the cryptographic protocol and its blockchain implementation. Bitcoin blockchain allows individuals to share value, i.e., exchange cryptocurrencies, by maintaining a distributed ledger. This ledger can be considered a distributed database, maintained by blockchain nodes storing the list of all transactions from the beginning of the system, updated continuously as the transaction arrives.

The novelty of this system stems from a property of its cryptographic protocol: it works without the need for a trusted third party. No one is in charge of the system, which cannot be controlled by anyone —unless by controlling the majority of the network's computation power or total coin value. The system relies on miners competing at solving complex cryptographic puzzles to obtain a right to validate transactions, together with a financial reward composed of newly created coins. Unlike state-controlled or private digital currencies, no one can forbid transactions from or to certain entities or reverse transactions due to their decentralized nature.

\section{Cryptocurrencies as Socio-Technical Systems}

What turned this technical innovation into a global phenomenon was the progressive adoption of Bitcoin by millions of individuals worldwide. As more people joined, an ecosystem of services and tools started to appear. Companies began to sell goods and services, such as the infamous Silk Road, known for selling drugs and engaging in criminal activities. Online casinos allow users to gamble without any legal entity's control. Other companies started a business of mining coins using ever-more performant hardware. States facing challenges with their own currencies, such as El Salvador, proposed to use Bitcoin as the official currency. Companies behaving like retail banks, known as exchanges, quickly became central to this ecosystem. They allowed customers to exchange Bitcoins for US Dollars and other fiat currencies. 
Following Bitcoin's success, many other cryptocurrencies have been created. Ethereum is the most famous public blockchain. Vitalik Buterin, a cryptocurrency researcher and programmer, proposed Ethereum in late 2013. It went live in 2015 with an initial supply of 72 million coins. Ethereum introduced the concept of smart contracts—self-executing contracts where the terms of the agreement are directly written into code \cite{wood2014ethereum}. These allow users to build more complex elements in the ecosystem, including fungible and non-fungible tokens (NFTs) and decentralized finance (DeFi) services, allowing not only to share virtual coins but also to lend, borrow or exchange money without a trusted third party, interacting directly with blockchain-based services.
Nowadays, cryptocurrencies are much more than simple digital assets used by a few traders for investments. They represent a whole ecosystem involving millions of individuals performing transactions among themselves and with large-scale companies, involving billions of US dollars in trading volumes every day.

\section{Cryptocurrency Transaction Analysis on Bitcoin}
Bitcoin was the first modern cryptocurrency and is also one of the simplest. Unlike later ones, such as Ethereum, Bitcoin is mostly limited to financial transactions and does not allow complex smart contracts; it can be mostly seen as a simple collection of financial transactions.
Bitcoin transactions can be studied as a network from two perspectives. The first consists of directly studying the transactions between Bitcoin addresses as they are stored in the blockchain. However, due to the way the Bitcoin protocol works, this network is hard to interpret: most users have multiple addresses, many addresses are used only once, and the network is not a simple directed graph, but a directed hypergraph: each Bitcoin transaction can involve multiple addresses as input and output. For this reason, the transaction network is often studied in a transformed form, a user transaction network.

\subsection{Bitcoin transaction networks
}
\begin{figure*}[h] 
    \centering
    \includegraphics[width=0.8\textwidth]{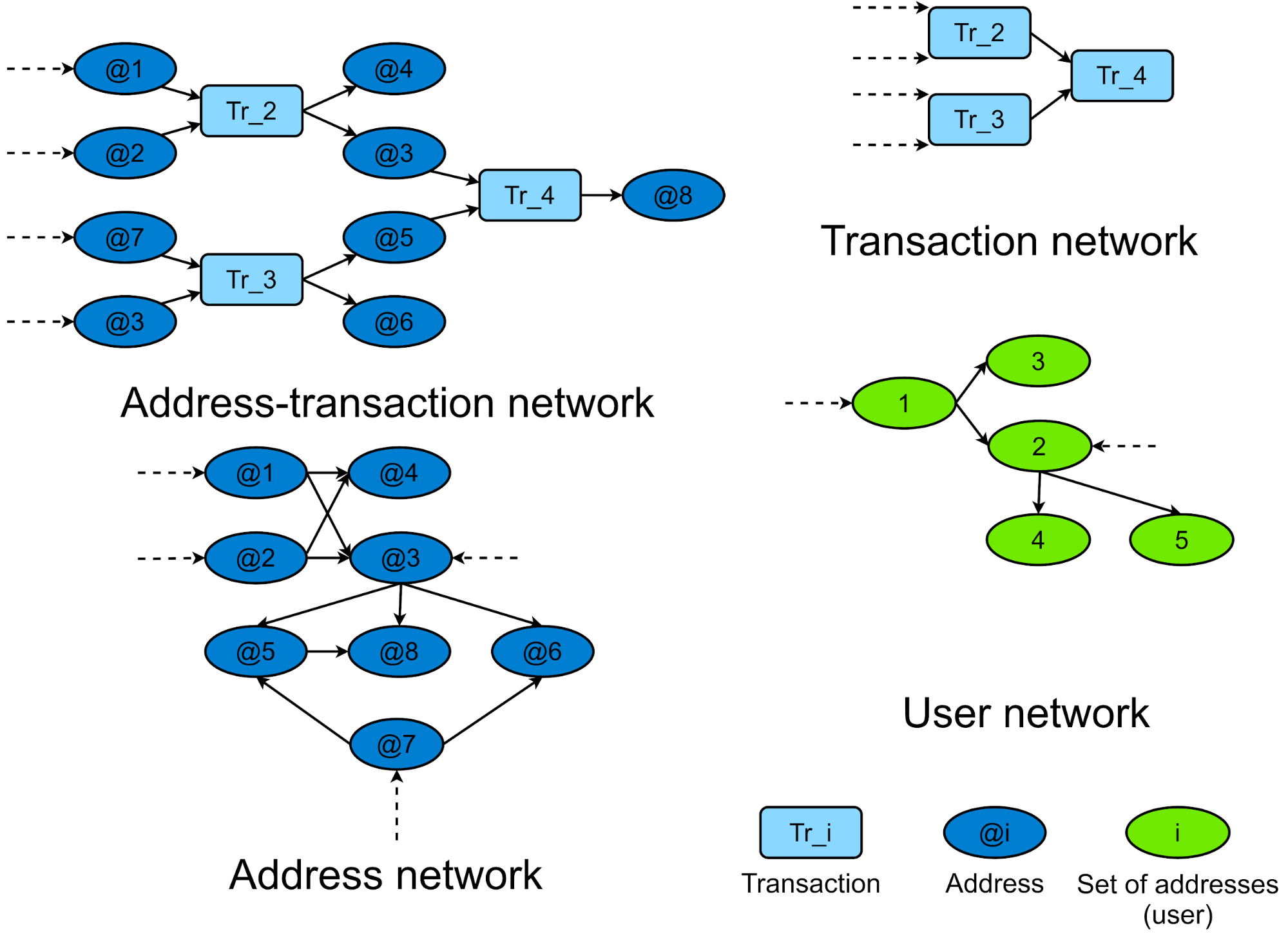} 
    \caption{Bitcoin transaction network representations.}
    \label{fig:bitTransNet} 
\end{figure*}

Each Bitcoin transaction, as stored in the blockchain, has n inputs and m outputs, each input contributing a unique value and each output receiving a unique value. The sum of the outputs is equal to the sum of the input minus the transaction fees. Multiple inputs are part of the nature of the Bitcoin protocol, which is known as UTXO (Unspent Transaction Output). According to this model, users do not store their digital coins in a unique account like a traditional bank account. Instead, every time they receive coins from a transaction, the corresponding output of the transaction is “secured” by their public key–also known as the Bitcoin address. The owner of this public key can then spend the coins in this particular output by signing–proving that they own that public key–using their private key. Thus, at any point in time, an active user typically controls multiple outputs of multiple transactions. The sum of all those controlled outputs corresponds to the amount of coins that they own. To improve users’ anonymity, it is recommended—since Nakamoto’s white paper—to avoid reusing multiple times the same public keys —i.e., Bitcoin addresses. Although many Bitcoin users reuse those addresses for convenience (for instance, using a paper wallet with the address as a QR code or requesting payment by posting their address in social media or internet forums), in general, one can assume that each user controls multiple addresses. Thus, transactions between addresses cannot be considered a good proxy for transactions between users, i.e., the most informative data sought for analyzing Bitcoin activity. 
From this raw data present in the blockchain, several network representations are possible, and researchers have used them for different purposes (Fig. \ref{fig:bitTransNet}).

\emph{Address-transaction network.}  It is a bipartite network representation, with nodes representing transactions and addresses. Directed edges represent the presence of addresses in the input and output of a transaction, weighted by the value they receive or contribute. This representation is without loss of information but is difficult to study due to its bipartite nature.

\emph{Transaction network.} One can create a transaction network by projecting the bipartite address-transaction network on transactions, considering that an edge exists between transactions Tx1 and Tx2 if there is at least one output of Tx1, which is also an input of Tx2. The edge is typically weighted by the sum of all such occurrences. This representation focuses on the flow of coins and does not represent users.

\emph{Address network.}
Conversely, one can also project the address-transaction network on addresses, thus considering that there is a directed edge between addresses @1 and @2 if @1 appears as input of a transaction and @2 in output of that same transaction. If the temporal aspect is ignored, a single edge can summarize multiple transactions. A weakness of this representation is that, due to the multiplicity, each transaction can lead to a large number of edges, particularly when considering large transactions made by companies that can involve hundreds of addresses as input and output. A more accurate representation would use a directed weighted hypergraph or higher-order network. However, this representation is more complicated to manipulate.

\emph{User network.} The most commonly used representation of cryptocurrency analysis consists of building a surrogate network, requiring a non-trivial process called address clustering, in which nodes correspond to users, i.e., groups of addresses, and edges correspond to transactions between these users. This representation also permits thinking in terms of Bitcoin balance for each user, i.e., the amount controlled at any given time by a user, defined as the difference between spent and received Bitcoins. The address clustering process is described in the next section. Although much more straightforward to interpret, this representation is not entirely faithful to what genuinely occurred in the blockchain for two reasons: 1) The details of through which transaction the money went from one user to the next is lost, and 2) The address clustering process is based on assumptions that are known to be imperfect, introducing inaccuracies in the resulting network.

\subsection{Research questions and key findings}

\begin{figure*}[h] 
    \centering
    \includegraphics[width=0.9\textwidth]{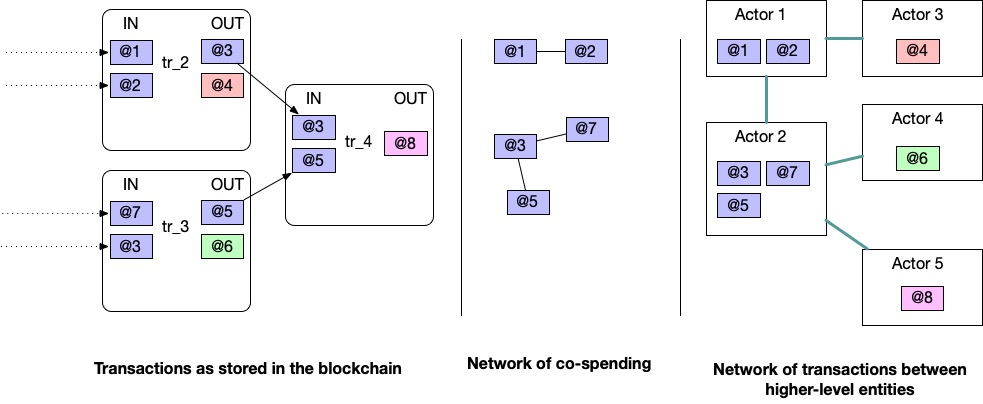} 
    \caption{Illustration of the address clustering approach based on the multi-input heuristic.}
    \label{fig:addressClustering} 
\end{figure*}

\emph{Address clustering.} The user or entity network is built by aggregating sets—clusters—of addresses belonging to the same entity, which can be an individual, a company, etc, as long as this entity controls a pool of addresses they manage conjointly. Address clustering techniques are diversified. The most common method, the co-input or co-spending heuristic, aggregates multiple addresses used as inputs of the same transaction (Fig. \ref{fig:addressClustering}) \citep{Reid2013,Ron2013}. It is well justified by the fact that having multiple entities signing a transaction together requires a high degree of cooperation and trust between them. This heuristic has been continuously shown to be reliable and effective \citep{Harrigan2016}. However, it is known to lead to an underestimation of the user’s set of addresses, for instance, because users voluntarily use hiding schemes such as fork-merge and peeling chain patterns \citep{Ron2013}. Another type of error, much rarer, occurs when users decide to merge their owned transaction output as input for a common transaction (coin merge) \citep{Ron2013}. To improve the address clustering process, multiple solutions have been proposed to discover so-called change addresses, such as more advanced heuristics \citep{Meiklejohn2013}, unsupervised machine learning \citep{Cazabet2017}, and supervised machine learning \citep{moser2022resurrecting,tubino2022towards}

\begin{figure*}[h] 
    \centering
    \includegraphics[width=0.8\textwidth]{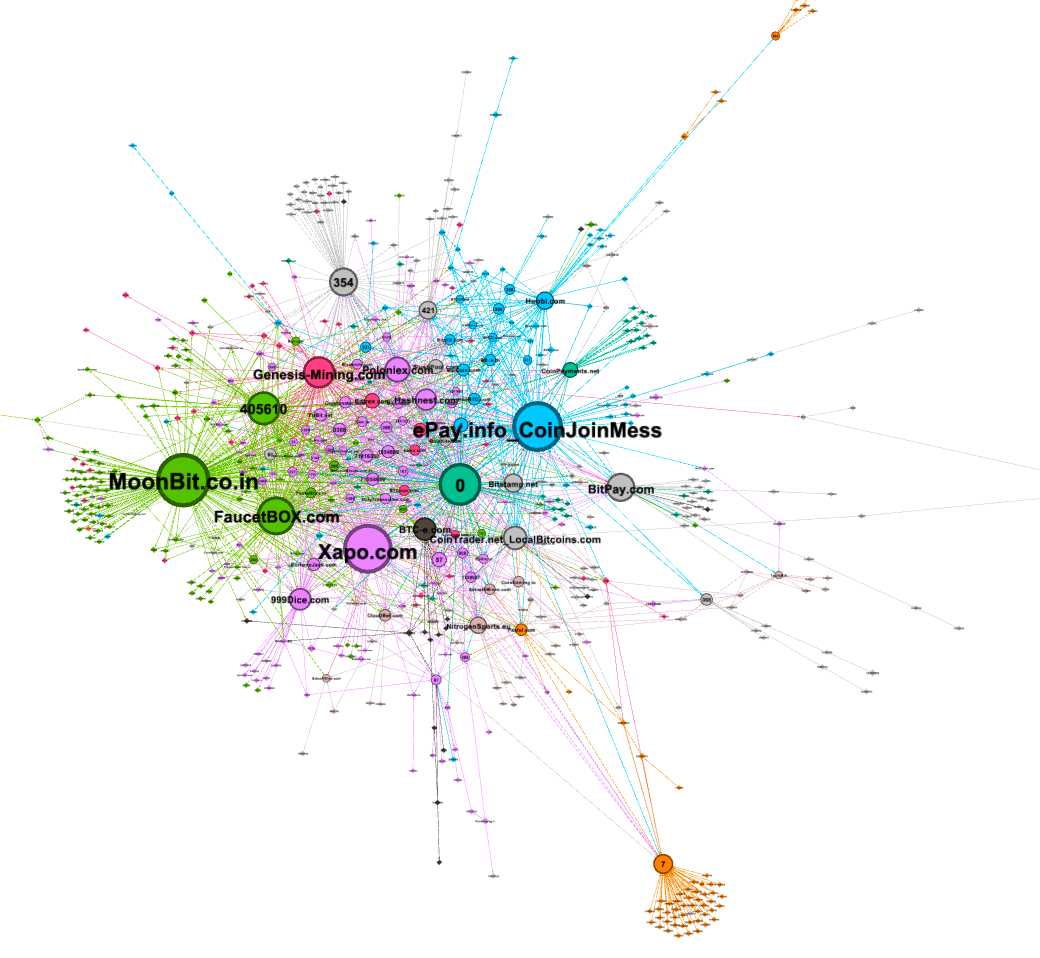} 
    \caption{Visualization of the Bitcoin user network.}
    \label{fig:BTCnet} 
\end{figure*}

\emph{Entity deanonymization and classification.}
Once address clusters have been identified, the analysis can be made richer by integrating information on the users. The most straightforward approach consists of using out-of-chain information: Web pages provide lists of important Bitcoin actors and their corresponding addresses (e.g., WalletExplorer or repositories associated with criminal activities), while other users can be found by scraping specialized forums and social media. An illustration of such a user network with labeled entities is shown in Fig. \ref{fig:BTCnet}. More complex techniques monitor the peer-to-peer network to match IP addresses with users \citep{zhu2017mining}. Beyond simply associating a name to an address cluster, many works, e.g., \citep{Jourdan2018} use machine learning techniques to infer the probable nature of unknown entities based on examples taken from external sources. Typical categories to infer are Gambling services, Mining pools, Exchange platforms, criminals, etc. Finally, some methods search to identify multiple separate accounts for the same user, even without direct interactions between them, by looking for similar behavioral patterns, such as \citep{monaco2015identifying} with address-address transactions time series and \citep{Tovanich2023}, using tainted flow analysis to build users fingerprints.

\emph{Structure and dynamics of Bitcoin user networks.}
A popular research question involves studying the properties of the transaction network and its evolution \citep{Maesa2018}. Global network characteristics such as the small world and scale-free properties \citep{Baumann2014} have been demonstrated for Bitcoin user networks. The well-known bow-tie structure of the World Wide Web is also a characteristic of this network \citep{Maesa2019}. In that study, the authors also found that the network's strongly connected component includes most of the transactions and that most miners are located in its incoming (IN) component.
In the context of economic perturbation and its effects on the network, \citep{Coquide2019} proposed a contagion model showing that the system's resilience is fragile once the most influential nodes are considered to be bankrupt. 

\emph{Profiling Bitcoin usages and users.} 
A question frequently asked by non-specialists concerns the usage of Bitcoin: Are most transactions due to financial trading and illegal activities?, or Is there a real usage for buying goods and services? We can be certain that blockchain data is not due to trading activities since exchanging Bitcoin against fiat currency is done through Exchange platforms by private scripture, without any activity in the blockchain. The transaction costs and technical constraints also make it unlikely that Bitcoin is used massively for daily payments. Several works focused on giving more details on these usages. For instance, \citep{Meiklejohn2013} found that a large fraction of transactions were due to gambling websites at some point in the early period. \citep{ramos2023temporal} distinguished between transactions made for technical reasons and genuine transactions between users, as well as identifying transactions related to commercial activities. They also proposed a method to infer geographic locations based on the time zones of the transactions. Other works focus, for instance, on the centralization trend in Bitcoin, with exchange platforms acting as banks and concentrating a large fraction of the total activity.

\emph{Economic properties.}
The distribution of wealth among Bitcoin users has been studied in several works. For example, a large amount of Bitcoin is kept by a few users \citep{Ron2013}. A rich-get-richer phenomenon is present, concentrating wealth with time \citep{kondor2014rich}. Numerous works have proposed methods to predict the evolution of the price of Bitcoin, sometimes based on fine-level analysis of the transaction network, including its topology, e.g., \citep{akcora2018forecasting}.

\emph{Fighting cybercrime.} Bitcoin is well known as a medium for performing illegal activities, such as buying illegal products and services, money laundering, etc. Several works have focused on detecting this type of activity, for instance, by tracking transactions related to famous pirates or observing FBI seizures. Some authors used supervised learning to infer users and transactions related to ransomware. A survey on this kind of work can be found in \citep{VanWegberg2018}.

\section{Smart Contract Analysis on Ethereum}
Ethereum is a public blockchain that implements smart contract functionality, called by its creators a “world computer.” A smart contract is a piece of program (code) that can be run on any machine in the Ethereum network through the Ethereum Virtual Machine (EVM). Ethereum ensures that smart contracts are Turing-complete by imposing so-called gas fees, representing a cost for every operation (e.g., computation and data access). Smart contracts enable new capabilities on the blockchain beyond only keeping records of native cryptocurrency.

\emph{Account-based model} Ethereum uses an account-based transaction model to keep track of the Ether balance (ETH, the Ethereum cryptocurrency unit). This mechanism is different from the UTXO used by Bitcoin and simplifies analysis since it is not mandatory to perform address clustering as a preprocess. Ethereum can thus be seen as a state machine, each node storing and synchronizing the current state of the blockchain. Each account's state, including its balance and stored data, is updated sequentially with the validation of each new block.
There are two different types of accounts in Ethereum. Accounts controlled by users are called externally owned accounts (EOA). Contract accounts (CA) store the smart contract code deployed by an EOA on the network. Both account types can receive, hold, and send ETH and other tokens, as well as call functions within smart contracts.

\emph{Tokens.} At the core of decentralized finance lies the emergence of digital assets on the Ethereum blockchain. This is realized through the implementation of smart contracts, which represent these assets—commonly referred to as tokens. These assets can be minted, burned, and exchanged while the contract account (CA) keeps track of the balance state for each account in its storage.

\vspace{0.5cm}

An asset can be fungible or non-fungible. Fungible assets (ERC-20 tokens) are interchangeable, and each unit is identical to another. They are typically used to create cryptocurrencies, such as Tether (USDT) or Wrapped Bitcoin (WBTC), that use the Ethereum blockchain as a protocol layer. They are also used as so-called governance tokens for DeFi smart contract services such as MakerDAO (DAO), Uniswap (UNI), and Compound (COMP). Non-fungible tokens (NFT), like ERC-721 tokens, are unique and cannot be exchanged on a one-to-one basis. Typical examples are digital collectibles or unique art pieces. Token standards (e.g., ERC-20, ERC-721) specify basic interfaces and allow third-party contracts to use tokens in a standardized way. These standards ensure interoperability and compatibility across the Ethereum ecosystem, facilitating a wide range of decentralized applications and services. All these tokens can be exchanged one against another, creating a network between currencies, smart contract services, and other types of digital assets.

\emph{Decentralized finance (DeFi).} 
 It refers to an alternative financial ecosystem using smart contracts to create financial services that do not rely on traditional intermediaries (e.g., banks, brokers, or governmental agencies). DeFi encompasses various protocols, such as decentralized exchanges, lending platforms, derivatives, and asset management, aiming to enhance efficiency, accessibility, and transparency in finance \citep{schar2021decentralized}. These protocols are controlled by predefined mechanisms in smart contracts, with policies set by the consensus of governance token holders. DeFi activities constitute a system on their own, in which one can study financial phenomena such as the emergence of price consensus, market mechanisms, and trading strategies.

\subsection{Ethereum network models}

\begin{figure*}[h] 
    \centering
    \includegraphics[width=0.8\textwidth]{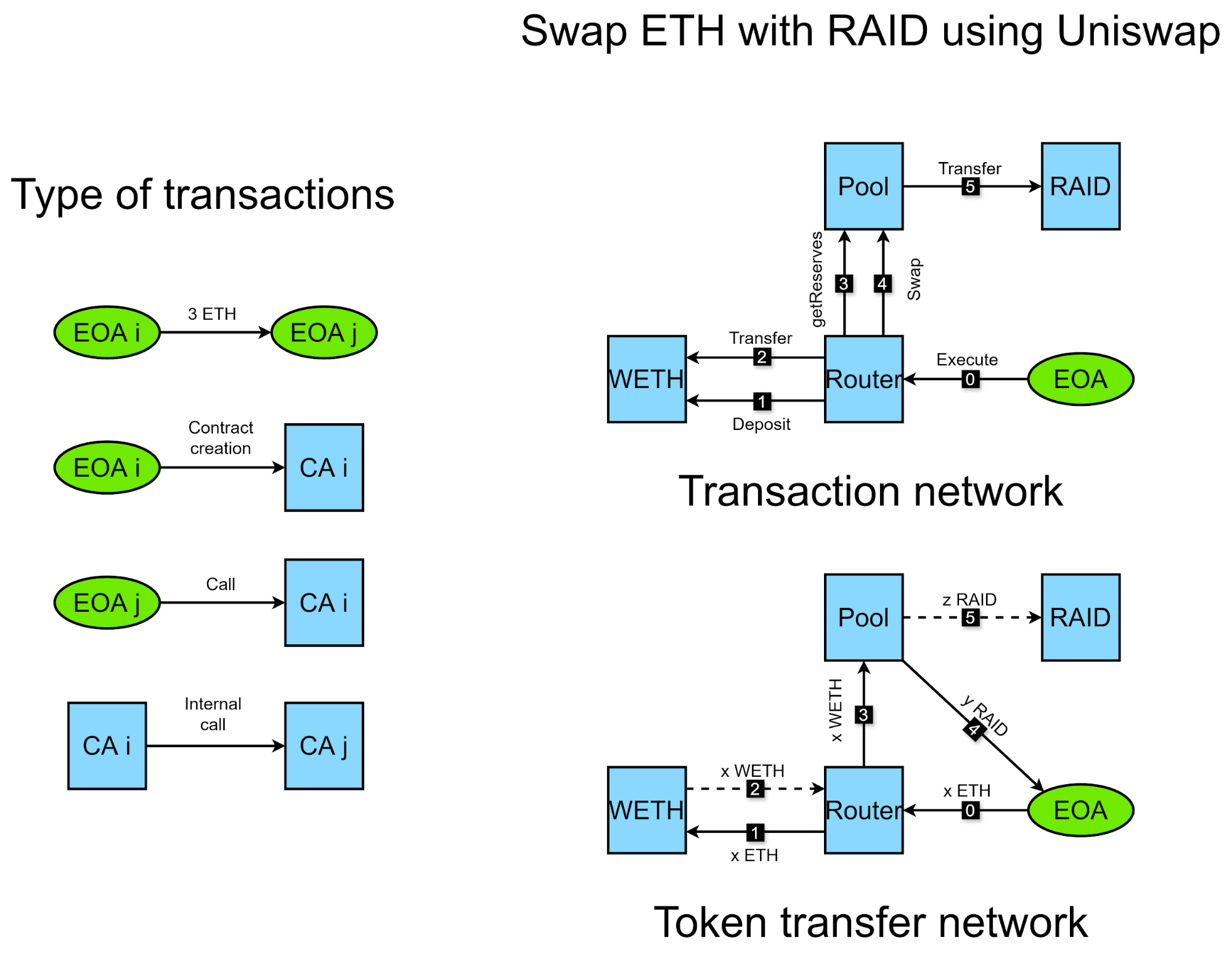} 
    \caption{Ethereum network representations.}
    \label{fig:etherNet} 
\end{figure*}

\emph{Ethereum transaction network} can be defined as a heterogeneous (multi-type) directed network (Fig. \ref{fig:etherNet}). Unlike the Bitcoin address network, there are three node types: EOA, contract accounts, and null. Edges can represent multiple things: ETH value transfers, smart contract creation, or function calls. The null address is a special account used to create and destroy smart contracts, as well as mint (create) and burn (destroy) tokens. This network represents all data stored in the blockchain, with all its richness and complexity.

\emph{Token transfer network} A network of token transfer between accounts can be constructed from a derivative of the Ethereum transaction network called the event logs, a record of events occurring during the execution of smart contracts in the Ethereum Virtual Machine (EVM).  Most ERC-20 tokens emit event logs for token transfers containing details about the sender, the receiver, and the amount transferred. Token transfer network represents the exchange of tokens between accounts. The directed and weighted edges represent the amount of tokens transferred from one entity to another. As accounts can exchange multiple tokens, there can be multiple edges between addresses for different tokens. This network can be analyzed in a way similar to what we introduced for the Bitcoin user network.

\subsection{Research questions and key findings
}

\emph{Structure and dynamics of Ethereum transaction networks.} As with Bitcoin, one can start by analyzing the network properties of different types of Ethereum networks. \citep{khan2022graph} provides a comprehensive survey of past work based on the network constructed and the analysis of its properties. Of particular interest is the evolution of the Ethereum network, which can be analyzed from the temporal properties of snapshot networks.

\emph{Detecting cybercrime in Ethereum} relies on labeled datasets of illicit accounts or activities. By tracking the money flow from illicit accounts, we can profile those accounts using network features and graph motifs. The results show that money laundering accounts have different characteristics from normal ones \citep{lin2023towards}. Furthermore, graph machine learning methods, such as Graph Neural Networks (GNNs), have been proven to be effective in classifying phishing scam accounts \citep{chen2020phishing}.

\emph{Non-fungible tokens (NFT).} Since its market boom in 2021, NFT has gained widespread public interest. They serve as proof of ownership for unique items like digital arts, music, or metaverse goods and items. NFT transaction network analysis reveals that interactions between traders follow the power law common in social and token networks \citep{casale2021networks}. In the NFT market, wash trading occurs when traders create fake transactions to inflate the perceived value of an NFT. These can be detected through instances of cyclical patterns conducted in a rapid sequence \citep{von2022nft}.

\emph{DeFi trading strategies.} While most of the trading activity occurs—as with Bitcoin—in private exchange platforms, thus inaccessible to analysts, DeFi allows users to engage in financial activities —typically, coin swaps, lending, borrowing, etc.— on the blockchain, thus letting public traces that can be analyzed \citep{schar2021decentralized}. For instance, \citep{kitzler2023disentangling} analyzed the topology of CA networks in 23 protocols and proposed an algorithm to extract the frequent DeFi building blocks. They reported that the most common building block is token swaps. Systemic risk can also occur in the DeFi ecosystem. This can be assessed by constructing the financial network and simulating which pools or entities in the protocol are more likely to propagate a domino effect \citep{tovanich2023contagion}.

\emph{Maximal Extractable Value (MEV)} refers to the strategy that block builders use to select and reorder transactions to optimize their profitability. EigenPhi is a blockchain explorer website that develops the heuristics to identify various types of MEV transactions. One common strategy is arbitrage, which exploits profits from price differences across different DeFi protocols. These opportunities can be detected from cycles in token transfer graphs \citep{zhou2021just}. A GNN-based model has been deployed to classify different types of MEV strategies, demonstrating more effective than conventional heuristic algorithms \citep{park2024unraveling}.

\section{Data Accessibility for Public Blockchains
}
A wonderful opportunity offered by cryptocurrencies is that all the data stored in their blockchain is accessible by anyone, thanks to their decentralized nature. Researchers thus have the opportunity to access this data without cost or limitations. Unlike private social networks such as Facebook or Twitter, there is no need to request authorization and no possibility from an owner to restrict access to this information. The details of the stored data depend on the cryptocurrency, but in most cases, one can at least access the sender and receiver addresses, amount, timestamps, and fees paid.

\emph{Running your own node.} Transaction data for the whole blockchain can be directly downloaded by running a full node. In the case of Bitcoin, transforming raw data into a network can be performed by following some simple steps, such as using ETL tools. 

\emph{ETL (extract-transform-load) tools.} These tools permit the extraction and transformation of all blockchain data or a part of it, leading to the facilitation of large blockchain data analysis. For example, Blockchain ETL provides open-source ETL scripts for various blockchains, including Bitcoin and Ethereum (\url{https://github.com/blockchain-etl}). These databases can be accessed on the Google BigQuey platform. Cryo provides a command line interface (CLI) and Python library for Ethereum ETL (\url{https://github.com/paradigmxyz/cryo}), offering a framework for efficiently accessing and processing large amounts of blockchain data from archive nodes.
\emph{Node-as-a-Service.} Several providers offer node-as-a-service solutions for those who prefer not to run their own node. These services, such as Infura, QuickNode, Moralis, and Alchemy, provide access to blockchain data without the overhead of maintaining the infrastructure. Some of them also offer their own APIs to interact with the blockchain and retrieve specific data like transaction details, token metadata, and NFTs.

\emph{Access transaction data through APIs. }Transaction data can be accessed more limitedly through APIs, such as Blockchain.com (Bitcoin), Etherscan (Ethereum), Blockchair, Bitquery, and Covalent (multiple blockchains). Some impose query limitations and may require payment for access beyond certain usage thresholds.

Understanding the specific protocols, functions, and emitted events is crucial for analysts interested in DeFi data. Tools like The Graph and Dune Analytics can be particularly useful in this regard. The Graph is a decentralized protocol for indexing and querying data extracted from DeFi protocols. Dune Analytics offers an SQL query engine and dashboard-building tool for raw blockchain data and extracted DeFi protocols.

\emph{Address taggings.} Due to the pseudonymous nature of blockchain users, blockchain intelligence companies such as Elliptic and Chainalysis work on deanonymizing these addresses and offer address tagging datasets, typically for a fee. WalletExplorer provides a list of addresses belonging to some well-known entities in Bitcoin, although its data was last updated in 2016. Etherscan, a widely used Ethereum explorer, also features labeled addresses, such as those belonging to exchanges or known protocols.

\emph{Curated datasets}, such as those from BADX, Chartalist, Elliptic dataset, and XBlock, provide specific benchmark data, e.g., on ransomware, phishing accounts, and token networks. These datasets can serve as a starting point for researchers, offering pre-processed data that can be used to validate new models or approaches.

\section{Key Applications
}
The first key application of cryptocurrency network analysis is related to finance, in particular, understanding the evolution of cryptocurrency prices and the probability of observing a crash or another major, sudden change in value.
Another key application concerns forensics: Many illegal activities occur on cryptocurrency and can be related to a particular entity. Tracking the movements of such illegal amounts of money and/or reidentifying the sources of entities trying to conceal this type of activity requires advanced mining of transactional data.
Finally, a last key application is simply understanding the evolution of the Bitcoin ecosystem as a socio-technical system, i.e., understanding emergent properties, activities, and behaviors in this system.

\section{Future Directions
}
Cryptocurrencies are evolving at a fast rate, adapting to societal and technical challenges. New questions are thus continually emerging, much as new phenomena such as Smart Contracts, NFTs, and DeFi emerged in the past. Recent trends such as SocialFi (Social Finance, enabling users to monetize their social media contents through cryptocurrencies) and perspectives such as the integration of cryptocurrencies into the Metaverse are possible new directions of research.  Among existing tasks, a future research direction is to use advanced neural network-based approaches, with potential interest for Bitcoin address clustering, forensics, user type classification, etc. Graph neural networks (GNNs) are particularly adapted to the study of transactional data, although they remain in little use due to the difficulty of scaling to very large datasets.

\bibliographystyle{plainnat}
\bibliography{ref.bib}

\end{document}